\documentclass[12pt, a4paper]{article}
\usepackage[utf8]{inputenc}
\usepackage{dcolumn,lscape}
\usepackage{amsmath,longtable,multicol,dcolumn,tabularx,amssymb}
\usepackage{exscale,amsthm,multirow,rotating}
\usepackage{natbib}
\usepackage{bbm}
\usepackage{tikz,dsfont,float}

\usepackage{todonotes}

\usepackage{hyperref}

\hypersetup{
    colorlinks=true,
    linkcolor=blue,
    citecolor=blue,
    filecolor=magenta,      
    urlcolor=cyan,
    pdftitle={review paper},
    pdfpagemode=FullScreen,
}

\usepackage{graphicx}
\usepackage{caption}
\usepackage{subcaption}

\usepackage{changes}
\definechangesauthor[name={PM}, color=blue]{PM}
\definechangesauthor[name={MP}, color=red]{PO}
\definechangesauthor[name={AF}, color=violet]{AF}

\graphicspath{{"image/"}}

\usepackage{algorithm}%
\usepackage{algorithmicx}%
\usepackage{algpseudocode}%

\begin{document}

\def\spacingset#1{\renewcommand{\baselinestretch}%
	{#1}\small\normalsize} \spacingset{1}


\title{\bf Multidimensional spatiotemporal clustering -- An application to environmental sustainability scores in Europe}
\author{Caterina Morelli\footnote{Corresponding author: c.morelli12@campus.unimib.it},  Simone Boccaletti,\\
	\small{Department of Economics, Management and Statistics (DEMS), University of Milano-Bicocca}\\
	Paolo Maranzano,\\
	\small{Department of Economics, Management and Statistics (DEMS), University of Milano-Bicocca, Italy} \\ \& \small{Fondazione Eni Enrico Mattei (FEEM), Italy}\\
	Philipp Otto\\
	\small{School of Mathematics and Statistics, University of Glasgow}}
\maketitle

\begin{abstract}
	The assessment of corporate sustainability performance is extremely relevant in facilitating the transition to a green and low-carbon intensity economy. However, companies located in different areas may be subject to different sustainability and environmental risks and policies. Henceforth, the main objective of this paper is to investigate the spatial and temporal pattern of the sustainability evaluations of European firms. We leverage on a large dataset containing information about companies’ sustainability performances, measured by MSCI ESG ratings, and geographical coordinates of firms in Western Europe between 2013 and 2023. By means of a modified version of the \cite{Chavent2018} hierarchical algorithm, we conduct a spatial clustering analysis, combining sustainability and spatial information, and a spatiotemporal clustering analysis, which combines the time dynamics of multiple sustainability features and spatial dissimilarities, to detect groups of firms with homogeneous sustainability performance. We are able to build cross-national and cross-industry clusters with remarkable differences in terms of sustainability scores. Among other results, in the spatio-temporal analysis, we observe a high degree of geographical overlap among clusters, indicating that the temporal dynamics in sustainability assessment are relevant within a multidimensional approach. Our findings help to capture the diversity of ESG ratings across Western Europe and may assist practitioners and policymakers in evaluating companies facing different sustainability-linked risks in different areas.
\end{abstract}

\noindent%
{\it Keywords:} Spatial hierarchical clustering, Spatiotemporal hierarchical clustering, Environmental Social Governance (ESG) ratings, Sustainability patterns.

\spacingset{1.45} 

\newpage

\section{Introduction}\label{Sec_Intro}
As the world is facing a path toward a more sustainable, greener, and less carbon intense economy, Environmental, Social and Governance (ESG) practices are becoming more and more relevant from the company perspectives in mitigating sustainability-linked risks. Since the Paris Agreement of 2015, great attention has been paid to corporate sustainability performance, especially on environmental aspects such as Greenhouse Gas (GHG) emission levels.

According to the World Health Organization (WHO), almost all of the global population breathes air that exceeds WHO guideline limits and contains high levels of pollutants. Moreover, air quality is closely linked to the earth’s climate and ecosystems globally. Many of the drivers of air pollution (i.e. combustion of fossil fuels) are also sources of greenhouse gas emissions. Policies to reduce air pollution, therefore, offer a win-win strategy for both climate and health, lowering the burden of disease attributable to air pollution as well as contributing to the mitigation of climate change.

In this context, firms contribute significantly to the emission of polluting gases with respect to households.
In Germany, the Environment Agency of the German Government \cite{uba2024} shows that companies produced more than 85\% of the $CO_2$ emissions in the last years. In the UK, the Department for Energy Security and Net Zero \citep{desnz2023} affirms that in 2022, household emissions account only for 17\% of the total. Moreover, the Italian National Institute of Statistics \citep{istat2022} shows that in the last 20 years, the overall volume of $CO_2$ has decreased by around 30\%, but the emissions from companies still represent around 70\% of the total.

Within this framework, Environmental, Social, and Governance issues have become crucial topics for companies' operations and stakeholders' engagement and activism. Companies have begun to make sustainability practices a core aspect of their operations and disclose more and more information on their commitment to environmental, social, and governance issues. Stakeholders have started to take ESG ratings and scores into consideration when making financial decisions, preferring to have relationships and interactions with companies that respect sustainability principles. ESG ratings and scores are synthetic evaluations from a specialized rater on a company’s sustainability performance from many different points of view, broadly grouped under the three pillars: E, S and G. As a result, empirical evidence highlights that companies that achieve good ESG ratings are able to obtain better financial conditions, e.g., lower cost of capital and easier access to capital markets.

In this paper, we aim to investigate the spatial pattern and the time dynamics of sustainability scores assigned to firms in Western Europe. We first trace the spatial pattern of Western European firm evaluation on three different sustainability levels, i.e., the overall sustainability-ESG score, the Environmental score, and the Carbon Emission performance 
using a tailored version of the hierarchical spatial clustering algorithm by \cite{Chavent2018}, which allows for detecting homogeneous groups of companies combining sustainability and spatial information. Furthermore, we examine the temporal dynamics of ESG-Environmental-Carbon emission evaluations in the last ten years by combining the spatial information and the similarity across the temporal series of multiple sustainability-related evaluations. The latter constitutes a multidimensional spatiotemporal extension of the spatial clustering methodology by \cite{Chavent2018}.

Our findings prove that both space and time dimensions are relevant in ESG performance evaluations. The initial spatial analysis, which is carried out for ESG ratings in 2023, provided evidence of the presence of cross-national and cross-industrial groups of companies with remarkable differences in the levels of environmental performance. Specifically, clusters are differentiated according to ESG scores, and the analysis brings out a cluster of companies with very poor sustainability performance, which belong to several European countries and are mainly classified in the manufacturing and mining industries. Other clusters are less transnational and are composed mainly of companies engaged in the tertiary and service sectors. Regarding spatiotemporal clustering, the identified groups are more prone to spatial overlapping, suggesting that the temporal aspect of the ESG scores is relevant to our multidimensional approach. 

The remainder of the paper is structured as follows. In Section \ref{Sec_Background}, we briefly review the current literature on ESG evaluation and patterns in the ESG scores. Then, we describe the dataset and the data collection procedure in Section \ref{Sec_Data_and_Methodology}. In Section \ref{sec:methods}, we introduce two multidimensional hierarchical clustering algorithms, the first for the purely spatial framework and the second for the spatiotemporal case. Thereby, we focus on techniques to efficiently select the hyper-parameters (i.e., weighting parameters and the number of clusters) by proposing two tailored algorithms. In Section \ref{Sec_Empirical findings}, we summarise the results of the cluster analyses and provide an interpretation and discussion of the identified clusters. Eventually, in Section \ref{Sec_Conclusions}, we sum up the main contents of the paper and provide concluding remarks and future research perspectives.

\section{Background}\label{Sec_Background}
The academic literature in the economic-financial field provides increasing evidence about factors that impact firms' ESG performance and the beneficial effects that ESG commitment has on companies and stakeholders. Given its relevant role, researchers have recently begun analysing ESG patterns to provide the basis for more specific studies on this phenomenon.


\subsection{The important role of ESG evaluation for firms and stakeholders}

First of all, companies' commitment to sustainability principles seems to be closely linked to environmental policies. In fact, \cite{Zhang2022} demonstrate that environmental regulation pushes companies to pay more attention to product quality and sustainability principles in production and \cite{Chen2022} find positive effects of environmental regulation on firm environmental investment. Also, \cite{Chen2022b, Qian2024, Xue2023}  observe a positive effect of green finance policy on ESG performance and \cite{Wang2022} show that green finance policy encourages enterprises to develop and adopt green products and technologies. Other research focuses on the Environmental Protection Tax Law in China, finding a positive effect on the ESG performance and green technological innovation \citep{Li2022}, particularly for heavily polluting firms \citep{He2023a, He2023b}. Moreover, \cite{Wu2023} show that executive green incentives and top management team characteristics positively impact the corporate ESG performance and \cite{Zhang2023} suggest that the disruption of environmental subsidies significantly positively affects them. Furthermore, the literature provides evidence of a positive impact of regional environmental transparency \citep{ Chen2023b}, digital transformation \citep{Zhao2023} and digital finance \citep{Mo2023} on ESG performance.

Among the evidence on the effects of ESG performance, \cite{Fu2023} found that it positively and significantly affects corporate financial performance, and digital transformation drives this promoting effect, \cite{Alfalih2023} show that social and governance dimensions of ESG influence companies’ financial performance across the two measures of a firm's financial performance (ROA and Tobin's Q), while environmental dimension is significant with the Tobin's Q measure. 
Also, \cite{Yu2022} find a significantly positive relationship between ESG composite performance and firm value and \cite{Panda2023} describe the positive effect of Corporate Sustainability expenditure on share prices. \cite{Lopez2023} demonstrate that the absence of CO2 equivalent emissions, the absence of incentives, and the presence of environmental investment have an impact on stock market returns. Moreover, the academic literature provides evidence of significant positive impact of ESG commitment on listed companies' stock liquidity \citep{Chen2023}, on productivity \citep{Ma2022}, on foreign investment flows \citep{Chipalkatti2021}, on green innovation \citep{Lian2023, Mukhtar2023, Zheng2023} and a negative impact on company over-indebtedness \citep{Lai2022}. 

The above-mentioned research highlights the relevance of ESG performance and, consequently, the fundamental role that disclosure of ESG assessments plays. 

\subsection{Pattern of firms' ESG evaluation}

The recent academic literature provides interesting examples of how researchers have traced and mapped different patterns of businesses of their ESG assessments throughout a cluster analysis or other methodologies for classification, sometimes considering specific aspects of sustainability or including variables related to specific features of the activities carried out by the companies themselves. 
In particular, \cite{Ronalter20239067} consider a sample of firms from Europe, East Asia and North America to perform a hierarchical cluster analysis including ESG indicators, and they carry out independence tests to compare the quality management systems and environmental management systems of firms with different ESG evaluation. \cite{Gonzaga2024} employed the Kohonen Self-Organizing Map for clustering developing market companies, providing valuable evidence of the changes in ESG scores over the course of the COVID-19 pandemic. Using the same methodology, \cite{Iamandi2019} examine the sustainability profile of European companies, considering the ESG score, the scores of the pillars and the scores of the indicators composing them.

Moreover, \cite{Wang2023} investigates the dynamics of three pillars of ESG scores among banks and observes a convergence of the evaluations in separate clusters in recent years, exploiting a specific panel data model proposed by \cite{ Phillips20091153, Phillips20071771} to represent the behavior of economies in transition, formulated as a nonlinear time-varying factor model. \cite{Saraswati2024112} use a sample of Indonesian firms and perform a K-means cluster analysis on ESG score pillars' score to clarify differences between ESG sustainability and practice and show the relationship among the distinct aspects of ESG performance. The authors identify three clusters using the Elbow method, the silhouette, and the Gap Statistical Method.

Focusing on the environmental aspect, \cite{Amores-Salvado2023431} consider a sample of public industrial firms from Europe, the United States and Canada, they classify them according to a four-position matrix based on the dichotomy environmental performance-disclosure, then they perform ANOVA tests showing differences in the groups according to nationality and sector.
\cite{Ishizaka2021} propose a new hierarchical multi-criteria clustering based on PROMETHEE. They take into account uncertainty and imprecision, making use of the Stochastic Multiobjective Acceptability Analysis (SMAA) and cluster ensemble methods, and they provide an interesting application on a sample of US banks, considering financial variables and ESG pillar scores.
\cite{Ortas2015673} exploit a multidimensional HJ-Biplot technique finding evidence of how different country-specific social and institutional schemes influence ESG evaluation in a sample of firms located in Spain, France and Japan.
\cite{Sariyer2022S180} consider a sample of companies listed in the (Borsa Istanbul) BIST sustainability index and, based on their ESG pillars scores, they perform a K-means++ algorithm which accounts for a smart centroid initialization method by assigning the first centroid randomly then selecting the rest of the centroids based on the maximum squared distance. Using the silhouette score, the authors identify heterogeneous clusters in terms of ESG evaluation and also in terms of size and profitability.

To the best of our knowledge, only \cite{Wang2023} explicitly account for the temporal component in the clustering algorithm. In the other research, in which data are observed in several years, the authors repeat the same analysis considering separately the different years and then interpret the evolution of clusters over time \citep{Ortas2015673,Gonzaga2024}. As regards the spatial component, although some studies have identified clusters with different compositions according to the country of origin of the companies \citep{Ronalter20239067,Ortas2015673,Amores-Salvado2023431}, but including spatial information into the clustering algorithm is still an open question.

The first innovation of our paper is the introduction of spatial and temporal components in tracing the patterns of companies' ESG evaluations. In particular, we want to include spatial and temporal information in the clustering algorithm so that companies belonging to the same group are geographically close and show a similar trend over time. This choice is motivated by two main reasons. First, as we have observed in the cited papers, sustainability commitment may vary based on the areas considered and, over time, according to different laws, the types of companies located in specific areas. Second, this study could be preliminary to other research aiming to compare firms' environmental commitments with respect to their observed environmental impacts. It considers the air quality in the area where firms are located, the presence of pollutants in the soil and water, and the production of waste, which are usually described by spatiotemporal models.

\section{Data collection and descriptive statistics}\label{Sec_Data_and_Methodology}

We use a unique dataset covering companies from 15 European countries. This dataset includes assessments of companies' sustainability performance and their geographical location.  Companies' ESG performance is measured through ESG ratings. We collect this information on the companies' ESG rating and its components from  MSCI ESG Ratings. ESG ratings are firm-level observations of their sustainability performance using different types of data, including, among others, sustainability reports, media sources, and specific surveys to the clients. MSCI provides ESG ratings for more than 10k companies at the worldwide level, resulting in an overall evaluation system\footnote{The assessments are not meant to be taken as absolute values, but need to be} with classes (i.e. AAA, AA, A, BBB, BB, B, CCC in order from best to worst), a numerical overall score that ranges between 0 (worst) and 10 (best). The overall score is a weighted average of the scores on the three main pillars (Environmental, Social, and Governance). MSCI also provides information on key issues under the three pillars, e.g., those relevant to our analysis, such as the carbon emission scores. The methodology that determines the aggregate pillar score from specific items, as well as the aggregation of the three pillars, E, S, and G, in the overall score and rating, are based on industry weights, reflecting the idea that valuation on key parameters is different according to the industry in which firms operate. As an example, the Environmental Pillar will weigh more on the final overall score for utilities than for firms in the Media \& Entertainment industry. Lastly, ESG scores mainly cover listed firms, which are the ones that are most frequently under the attention of society, investors, and policymakers regarding sustainability issues.

Henceforth, ESG ratings are not just climate ratings. If a company’s greenhouse gas emissions pose significant financial risks, its ESG rating will reflect that. For example, direct emissions pose a significant risk to power and steel companies, while emissions from their products after they have left the factory gate can pose a significant risk to automobile companies. However, for industries such as health care, the most financially relevant risks lie elsewhere, so emissions have less influence on a company’s rating. For this reason, since this research focuses on the environmental aspects of sustainability scores, we consider the overall ESG score, the Environmental Pillar score, and the Carbon Emission score, which are the three indicators in the set of sustainability scores that consistently draw attention from both scholars and professionals in the field. We take the annual observations of these three variables from 2013 to 2023, considering only companies with a weight of the Carbon Emission score greater than zero, that is, companies whose activity involves the emission of greenhouse gases.

We match the ESG rating database with Orbis BvD database to link companies' ESG scores to their location. Initially, we collect the address of the Registered Office and NACE sector classification of listed companies located in Western European countries (Austria, Belgium, Denmark, France, Germany, Gibraltar, Ireland, Italy, Luxembourg, Malta, Netherlands, Portugal, Spain, Switzerland and United Kingdom). We include active and inactive companies so that we do not exclude observations from companies that may have had an ESG rating in past years but have recently been the subject of mergers or acquisitions or have left the market. 

The focus on Western Europe is to avoid problems in translating companies' addresses from different alphabets since the algorithm could miss-locate firms in these cases. Henceforth, companies from Eastern Europe, Greece, and Scandinavian countries are excluded. 

\begin{figure} 
\includegraphics[width=1\linewidth]{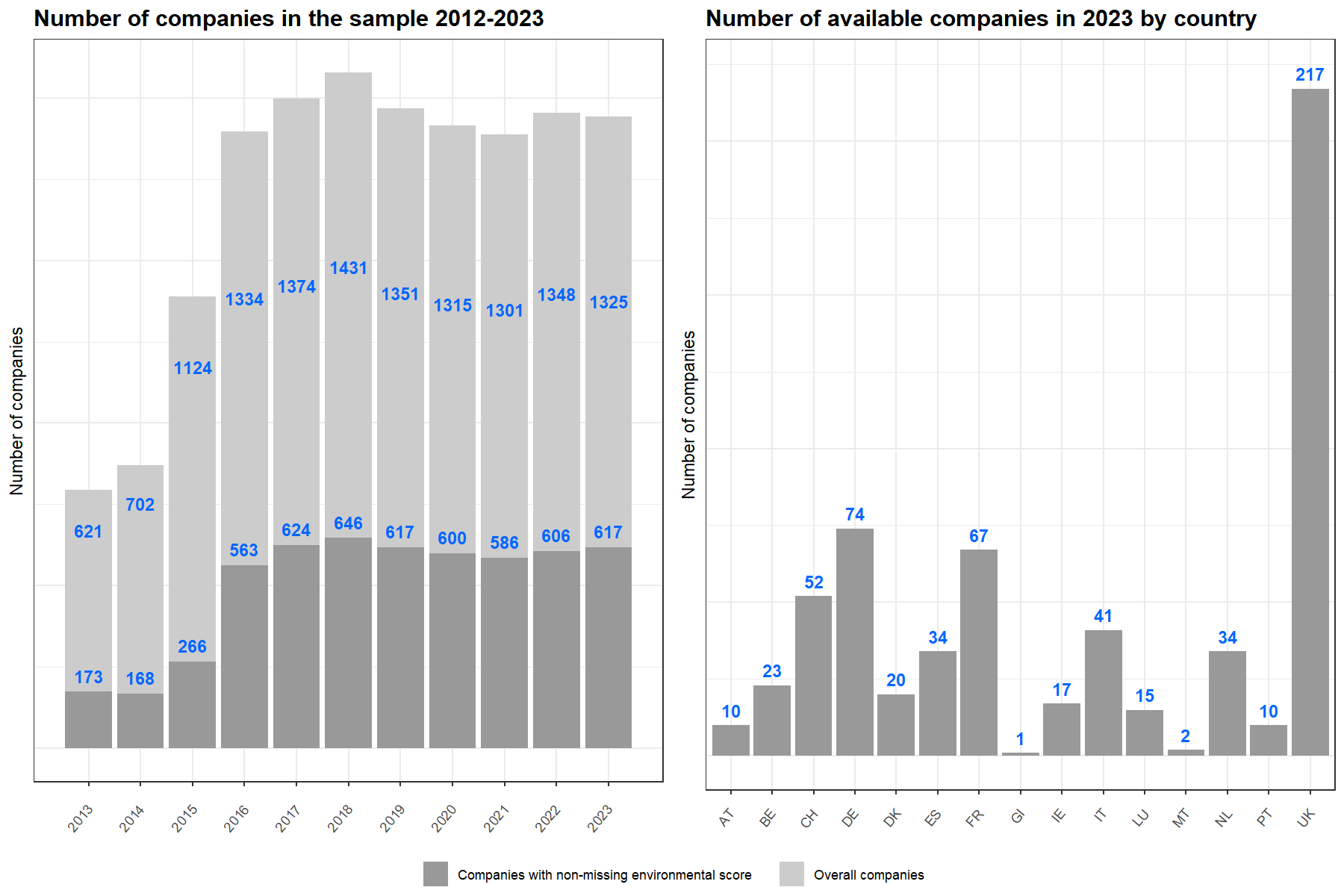}
\caption{Left panel: Number of observations per year between 2013 and 2023.  We report the total number of observations and the number of observations with a positive weight of the Carbon Emission score, which represents the observations included in our sample. Right panel: number of companies per country in 2023 with a positive Carbon Emission score weight}
\label{fig:SampleDescription}
\end{figure}

In Figure \ref{fig:SampleDescription} we provide an overview of the sample size, in particular considering the number of firms by year and the number of firms by country in the last year. In our spatial clustering, we are using only data from 2023, thereby considering 617 companies. In the spatiotemporal cluster analysis, we use a sample of 460 companies since we need companies with at least six observations in the time window considered, as described in the methodology section. We exclude observations from 2012 and earlier because of the small number of available ratings. 

\section{Methodology: hierarchical spatial and spatiotemporal clustering}\label{sec:methods}

The main goal of this paper is to investigate both the spatial and temporal patterns of the sustainability evaluations of Western European firms between 2013 and 2023. To do so, we carry out a spatial and spatiotemporal clustering analysis based on the methodology proposed by \cite{Chavent2018}, which combines socio-economic features, temporal dynamics and geographical information. Specifically, we implement a modified version of their Ward-like hierarchical algorithm \citep{Ward1963236} that, in addition to detecting homogeneous groups under geographical constraints, selects the clustering hyperparameters such that the total proportion of explained inertia is maximized. The algorithm is then employed in a spatial framework combining cross-sectional socio-economic features and spatial information and in a spatiotemporal framework leveraging on the time dynamics of multiple sustainability distances and spatial dissimilarities.

\subsection{Spatial hierarchical clustering}
Let $D = [d_{ij} ]_{i,j=1,\dots ,n}$ be the dissimilarity matrix of the observations and let $w_i$ be the weight of the $i$-th firm for $i = 1,\dots, n$. Without prior information, it is commonly set to $w_i=1/n$. Alternatively, the Ward hierarchical clustering approach starts with an initial partition in $n$ clusters of singletons, and at each step, the algorithm aggregates the two clusters such that the new partition has minimum within-cluster inertia. We define $\mathcal{P}_K=(\mathcal{C}_1, \dots, \mathcal{C}_K)$ a partition of the dataset into $K$ clusters and the pseudo-inertia of cluster $\mathcal{C}_K$ is computed as follows:
\begin{equation}
I(\mathcal{C}_K)=\sum_{i \in \mathcal{C}_K} \sum_{j \in \mathcal{C}_K} \frac{w_i w_j}{2 \sum_{i \in \mathcal{C}_K} w_i} {d_{ij}}^2.    
\end{equation}
The pseudo-within-cluster inertia of the partition is computed as the sum of the pseudo-inertia of each cluster. We point out that the pseudo-inertia is a generalization of the inertia when the dissimilarities can be non-Euclidean. From here on, we will always refer to pseudo-inertia, but for simplicity, we will call it inertia.

The spatial component is included by considering for the sample of $n$ units two $n \times n$ dissimilarity matrices, namely $D_0 = [d_{0,ij} ]_{i,j=1,\dots,n}$ and $D_1 = [d_{1,ij} ]_{i,j=1,\dots,n}$, referring to euclidean distances matrix of socio-economic variables under consideration and the geodetic distances matrix, respectively. Notice that since the distances in the $D_0$ and $D_1$ matrices may belong to two very different measurement scales (e.g., socioeconomic distance in currency and physical distances in kilometers), it is necessary to scale the dissimilarity matrices with respect to their maximum values so that the distances across observations take values between $0$ and $1$.

For a given mixing parameter $\alpha \in [0,1]$, it is possible to obtain a convex combination of the dissimilarity matrices $D(\alpha)=(1-\alpha)D_0+\alpha D_1$ and thus perform the hierarchical clustering algorithm. Notice that $\alpha$ states the importance of geographical and socio-economic information in determining the clusters. Indeed, as one set $\alpha = 0$, the geographical dissimilarities are not taken into account, while when $\alpha = 1$, the socio-economic distances are ignored, and the clusters are defined according to geographical distances only.

Given the partition $\mathcal{P}_K^{\alpha}=(\mathcal{C}_1^{\alpha}, \dots, \mathcal{C}_K^{\alpha})$, the mixed inertia for cluster $\mathcal{C}_K^{\alpha}$ is defined as the convex combination between the attribute inertia, and the inertia of the spatial component
\begin{equation}
I(\mathcal{C}_K^{\alpha})=(1-\alpha)\sum_{i \in \mathcal{C}_K} \sum_{j \in \mathcal{C}_K} \frac{w_i w_j}{2 \sum_{i \in \mathcal{C}_K} w_i} {d_{0,ij}}^2 + \alpha \sum_{i \in \mathcal{C}_K} \sum_{j \in \mathcal{C}_K} \frac{w_i w_j}{2 \sum_{i \in \mathcal{C}_K} w_i} {d_{1,ij}}^2    
\end{equation}
and the mixed within-clusters inertia is computed as the sum of the mixed inertia of its clusters, i.e.,
\begin{equation}
W_\alpha(\mathcal{P}_K^{\alpha})=\sum_{k=1}^{K} I_\alpha(\mathcal{C}_k^{\alpha}).    
\end{equation}

\subsection{Spatial hierarchical clustering: choice of the parameters}
The main issue in such a hierarchical clustering approach is the choice of the parameters $\alpha$ and $K$. \cite{Chavent2018} suggest setting a prior value for $K$ and then providing a criterion to choose $\alpha$ such that it allows to explain the same proportion of the dissimilarities from both matrices, with respect to the cases in which the clusters are obtained considering only the feature matrix or the spatial matrix. They introduce the notion of the proportion of the total pseudo inertia explained by partition $\mathcal{P}_K^{\alpha}$ in $K$ clusters as:
\begin{equation}
Q_{\beta}(\mathcal{P}_K^{\alpha})=1-\frac{W_{\beta}(\mathcal{P}_K^{\alpha})}{W_{\beta}(\mathcal{P}_1)} \,   
\end{equation}
where $\beta$ can be either $D_0$ or $D_1$, depending on which dissimilarity matrix is used as a benchmark. Specifically, $Q_{D_{0}}(\mathcal{P}_K^{\alpha})$ quantifies the proportion of socio-economic pseudo inertia (i.e., $W_{D_0}(\mathcal{P}_1)$) explained by partition $\mathcal{P}_K^{\alpha}$, while $Q_{D_{1}}(\mathcal{P}_K^{\alpha})$ quantifies the amount of geographical pseudo inertia (i.e., $W_{D_1}(\mathcal{P}_1)$) explained by partition $\mathcal{P}_K^{\alpha}$.

To account for potential scale issues in $Q_{D_{0}}(\mathcal{P}_K^{\alpha})$ and $Q_{D_{1}}(\mathcal{P}_K^{\alpha})$, the $Q_{\beta}(\mathcal{P}_K^{\alpha})$ metrics are then normalized with respect to the baseline case of purely-geographical or purely-socio-economic clustering, that is, by computing the following ratios:
\begin{equation}
\tilde{Q}_{D_0}(\mathcal{P}_K^{\alpha})=\frac{Q_{D_0}(\mathcal{P}_K^{\alpha})}{Q_{D_0}(\mathcal{P}_K^0)} \qquad \tilde{Q}_{D_1}(\mathcal{P}_K^{\alpha})=\frac{Q_{D_1}(\mathcal{P}_K^{\alpha})}{Q_{D_1}(\mathcal{P}_K^1)}.
\end{equation}
This relative formulation allows for a straightforward interpretation of the values. For instance, by considering $\tilde{Q}_{D_0}(\mathcal{P}_K^{\alpha})$, for a given $K$ and a given mixing parameter $\alpha$, one is expressing the percentage improvement in the explained proportion of pseudo inertia obtained by using a mixture of geographical and socio-economic feature to generate the partition $\mathcal{P}_K^{\alpha}$ (i.e., $Q_{D_0}(\mathcal{P}_K^{\alpha})$) with respect to the proportion of pseudo inertia it would be explained by only using socio-economic feature to generate the partition $\mathcal{P}_K^{\alpha}$ in $K$ clusters (i.e., $Q_{D_0}(\mathcal{P}_K^0)$). Conversely, if one considers $\tilde{Q}_{D_1}(\mathcal{P}_K^{\alpha})$, the resulting value for a specific pair of $K$ and $\alpha$ expresses the improvement obtained by mixing the two dimensions instead of using a purely-geographical partitioning algorithm. Being $\alpha$ a measure of the trade-off between the loss of socio-economic homogeneity and the gain of geographic homogeneity, for a fixed $K$, increasingly values of $\alpha$ will correspond to higher $\tilde{Q}_{D_1}(\mathcal{P}_K^{\alpha})$ and lower $\tilde{Q}_{D_0}(\mathcal{P}_K^{\alpha})$. For a technical discussion about the properties of these quantities, we refer the readers to Section 3 in \cite{Chavent2018}.  

\cite{Chavent2018} suggest to choose $\alpha$ such that the normalized proportion of the explained pseudo inertia from $D_0$ and $D_1$ are as similar as possible, that is,
\begin{equation}
    min_{\alpha} |\tilde{Q}_{D_0}(\mathcal{P}_K^{\alpha}) - \tilde{Q}_{D_1}(\mathcal{P}_K^{\alpha})|
\end{equation}
which means to identify $\alpha$ such that socio-economic and geographical information return as similar as possible proportion of explained pseudo inertia with respect to proportion it would be explained by only using socio-economic features or spatial features to generate the partition. Consequently, the number of clusters can be chosen according to the dendrogram or elbow criteria. Following a similar rationale, \cite{Mattera2023} set an initial number of clusters $K_0$ considering the partition associated with $D_0$, then they determine $\alpha$ as in \cite{Chavent2018}, and finally they define the optimal number of clusters based on the combined dissimilarity matrix. Notice that this selection method does not always allow to identify $\alpha$ such that it captures the highest possible overall dissimilarity in the data. To address such drawback, \cite{Jaya2019} start finding $\alpha$ according to \cite{Mattera2023} while choosing a different mixing parameter in order to explain better the normalized proportion of inertia in one matrix, with a relatively small reduction of the normalized proportion of inertia from the other matrix.

Hereafter, we propose an algorithm to select the clustering hyperparameters, that is, the mixing coefficient $\alpha$ and the number of clusters $K$, that generalizes the aforementioned approaches by optimizing the \textit{weighted average of the explained mixed pseudo inertia}, which can be expressed in several ways: 
\begin{equation}
\begin{split}
    \bar{Q}(\mathcal{P}_K^{\alpha}) &= \frac{Q_{D_0}(\mathcal{P}_K^{\alpha}) \cdot W_{D_0}(\mathcal{P}_1) + Q_{D_1}(\mathcal{P}_K^{\alpha}) \cdot W_{D_1}(\mathcal{P}_1)}{W_{D_0}(\mathcal{P}_1) + W_{D_1}(\mathcal{P}_1)} \\
     &= \left[ 1 - \frac{W_{D_0}(\mathcal{P}_K^{\alpha}) + W_{D_1}(\mathcal{P}_K^{\alpha})}{W_{D_0}(\mathcal{P}_1) + W_{D_1}(\mathcal{P}_1)} \right].
\end{split}
\end{equation}
In particular, conditioning on a given $K$, the optimal $\alpha$ is given by the maximizer of $\bar{Q}(\mathcal{P}_K^{\alpha})$, that is, 
\begin{equation} \label{Eq_MaxCriteria}
    max_{\alpha} \bar{Q}(\mathcal{P}_K^{\alpha}).
\end{equation}
Thus, while \cite{Chavent2018} defined the optimal $\alpha$ as the one balancing the explained inertia from socio-economic and geographical features, we are proposing to select the $\alpha$, which jointly maximizes the amount of pseudo inertia explained from both the socio-economic and the geographical information (i.e., $Q_{D_0}(\mathcal{P}_K^{\alpha})$ and $Q_{D_1}(\mathcal{P}_K^{\alpha})$), weighted by the cumulated spatial and socio-economic pseudo inertia embedded the data (i.e., $W_{D_0}(\mathcal{P}_1) + W_{D_1}(\mathcal{P}_1)$). Further details about $\bar{Q}(\mathcal{P}_K^{\alpha})$, in particular, its relationship with the other metrics presented above are provided in Appendix \ref{App_DerivGain}.

The proposed algorithm works as follows. Having fixed a given value of $K$, the hierarchical clustering is performed for a sequence of $\alpha$ values on a regular grid from $0$ to $1$, with a specified constant increment $\Delta \alpha$. For any $K$, the best $\alpha^*_K$ is chosen to maximize the total proportion of explained inertia. The computation is iteratively repeated considering a range of $K$ up to a defined maximum, that is, $K=1,2,\dots,K_{max}$. In this way, we obtain an optimal weighting $\alpha_K^*$, conditional on $K$, which we then optimize across a range from $K = 1, ...K_{max}$. Then, the optimal number of clusters $K^*$ (evaluated at the optimal $\alpha^*_K$) is determined according to one or more suitable criteria for hierarchical clustering \citep{Kaufman1990}. For instance, consider computing the increments in the weighted average proportion of explained pseudo inertia (which can be interpreted as the increase of the explained variability) induced by a unitary increase in the number of groups or computing the values of the Silhouette index (synthesizing the average homogeneity of units within each cluster). According to the former, the optimal number of groups will match the largest value of $K$, guaranteeing a relevant increment in the weighted average explained inertia, while according to the latter, higher Silhouette values indicate that, on average, the units are properly matched within their own cluster. The proposed algorithm is summarized in Algorithm \ref{algorithm:sp}.

\begin{algorithm}
\caption{Hierarchical Spatial Clustering: grid for choice of $\alpha$ and $K$}\label{alg:cap}
\begin{algorithmic}
\State Define as $D_0 = [d_{0,ij} ]_{i,j=1,\dots,n}$ the feature dissimilarity matrix 
\State Define as $D_1 = [d_{1,ij} ]_{i,j=1,\dots,n}$ the spatial dissimilarity matrix  
\State Define as $K_{max}$ the maximum number of clusters
\State Define as $\Delta \alpha$ the increment of $\alpha$
\For {$K=1,\dots, K_{max}$}
    \For{$\alpha \in [0,1]$, by $\Delta \alpha$}
        \State Compute the linear combination of the two dissimilarity matrices $D(\alpha)=(1-\alpha)D_0+\alpha D_1$;
        \State Compute the $\mathcal{P}_K^{\alpha}=$ partition in $K$ clusters according to Ward hierarchical algorithm on the combined matrix $D$;
        \State Compute the weighted average of the explained mixed pseudo inertia $\bar{Q}(\mathcal{P}_K^{\alpha})$ 
    \EndFor
    \State Select the best $\alpha$ for each $K$ such that $\alpha_K^*=argmax_{\alpha} \bar{Q}(\mathcal{P}_K^{\alpha})$
\EndFor
\State Choose $K^*$ (evaluated at the corresponding $\alpha_{K}^*$) according to one or more hierarchical clustering criteria, such as the first difference in the weighted average proportion of explained pseudo inertia or the Silhouette index.
\end{algorithmic}
\label{algorithm:sp}
\end{algorithm}

\subsection{Spatiotemporal hierarchical clustering} \label{Sec_Method_STclu}
As regards the spatio-temporal clustering, several authors proposed to adapt the methodology proposed by \cite{Chavent2018} to the case of georeferenced time series data by combining the dissimilarity matrix computed on the $n$ time series and the spatial dissimilarity component.

At the best of our knowledge, even considering different fields of application, the literature considered time series related to only one socio-economic variable \citep{Bucci2023214, Deb2023, Mattera2023}. We aim to extend this framework by combining together multiple dissimilarities matrices corresponding to several time series of socio-economic features in addition to the spatial distances. Specifically, we combine the four dissimilarity matrices referring to the time series of the overall ESG score, the Environmental Pillar score and Carbon emission score, and the spatial component.

The distances among time series are computed adopting the Dynamic Time Warping (DTW) algorithm proposed by \cite{Zhang2006305} and implemented in the function \texttt{diss.DTWARP()} from \texttt{TSclust} package in \texttt{R} \cite{Montero20141}. DTW is a distance-minimizing temporal alignment between two time series that allows us to compute a dissimilarity measure among time series that could have different lengths and/or missing observations during the period but have at least one overlapping time stamp. Let us consider two time series, namely $x_t(t=t_{x_1},\dots , T_x)$ and $y_t(t=t_{y_1},\dots , T_y)$, such that $t_{x_1} \lesseqgtr t_{y_1}$ and $T_{x} \lesseqgtr T_{y} $ but $T_{x} \geq t_{y_1}$ and $T_{y} \geq t_{x_1}$. Let us compute the distance between two points $d(x_s,y_r)=|x_s-y_r|$ as the DTW distance between $x$ and $y$ up to points $s$ and $r$, which is given by the optimal alignment minimizing the following the distance:
$$
\Delta(s,r)=d(x_s,y_r)+\min [\Delta(s-1,r-1),\Delta(s-1,r),\Delta(s,r-1)].
$$

Once computed, the dissimilarity matrices $D_p = [d_{p,ij}]_{p=1,\dots, P; i,j=1,\dots,n}$, where $P-1$ is the number of variables included and $D_P$ refers to the spatial dissimilarity matrix obtained computing the geodetic distances across the observations, we can proceed in finding the parameters $K$ and $\alpha_p$ for the linear combination $D(\alpha_p)=\sum_{p=1}^P \alpha_pD_p$. Note that $\alpha_1,\dots,\alpha_{p-1}$ are the coefficients for the temporal dissimilarity matrices, while $\alpha_P=1-\sum_{p=1}^{P-1}\alpha_p$ represents the weight for the spatial component. We recall that dissimilarity matrices are normalized with respect to their maximum value.

We adapt the criterion proposed in Algorithm \ref{algorithm:sp} for the spatial clustering by choosing the vector $\alpha_p$ maximizing the weighted average of the explained mixed pseudo inertia induced by partition $\mathcal{P}_K^{\alpha_p}$.

Similarly to Algorithm \ref{algorithm:sp}, for a fixed number of groups $K$, we consider all the possible combinations for a grid of $\alpha_p(p=1,\dots,P)$ with a constant increase of each $\alpha_p$ equal to $\Delta \alpha$ and such that $\sum_{p=1}^P\alpha_p=1$. Then, we identify the clustering partition according to the Ward hierarchical algorithm on the combined distance matrix $D$. In particular, we adapt the criterion proposed in Algorithm \ref{algorithm:sp} for the spatial clustering by choosing the vector $\alpha_p$ maximizing the weighted average of the explained mixed pseudo inertia induced by partition $\mathcal{P}_K^{\alpha_p}$, given by the following generalization of the Equation \ref{Eq_MaxCriteria}:
\begin{equation}
\bar{Q}(\mathcal{P}_K^{\alpha_p})=1-\frac{\sum_{p=1}^P W_{D_p}(\mathcal{P}_K^{\alpha_p})}{\sum_{p=1}^P W_{D_p}(\mathcal{P}_1^{\alpha_p})}.    
\end{equation}
At this point, conditioning on $K$, we select the values of $\alpha_p$ for which $\bar{Q}(\mathcal{P}_K^{\alpha_p})$ is maximum. We iterate this step for a defined range of potential candidates $K=1,2,\dots,K_{max}$ evaluated at the corresponding optimal $\alpha_{K,p}^*$. Finally, we select the number of clusters according to the same criteria defined for the spatial clustering, that is, the increments in the weighted average proportion of explained inertia and the Silhouette index. The proposed algorithm is summarized in Algorithm \ref{algorithm:spt}.

\begin{algorithm}
\caption{Hierarchical Spatiotemporal Clustering: grid search of $\alpha_p(p=1,\dots,P)$ and $K$}\label{alg:cap}
\begin{algorithmic}
\State Define as $D_p = [d_{p,ij} ]_{p=1,\dots, P; i,j=1,\dots,n}$ the feature dissimilarity matrices
\State Define as $K_{max}$ the maximum number of clusters
\State Define as $\Delta \alpha$ the increment of $\alpha_p$
\For {$K=1,2,\dots, K_{max}$}
    \For{$\alpha_p(p=1,\dots,P) \in [0,1]$ by $\Delta \alpha$ such that $\sum_{p=1}^P\alpha_p=1$}
        \State Compute the linear combination of the dissimilarity matrices $D(\alpha_p)=\alpha_1D_1 + \alpha_2D_2 + \dots + \alpha_PD_P$
        \State Compute the partition $\mathcal{P}_K^{\alpha_p}$ in $K$ clusters according to Ward hierarchical algorithm on the combined matrix $D$
        \State Compute the weighted average of the explained mixed pseudo inertia $\bar{Q}(\mathcal{P}_K^{\alpha_p})$ for each partition
    \EndFor
    \State Select the best $\alpha_{Kp}$ for each $K$ such that $\alpha_{Kp}^*=argmax_{\alpha_p (p=1,\dots,P)}\bar{Q}(\mathcal{P}_K^{\alpha_p})$
\EndFor
\State Choose $K^*$ (evaluated at the corresponding $\alpha_{Kp}^*$) according to one or more hierarchical clustering criteria, such as the first difference in the weighted average proportion of explained pseudo inertia or the Silhouette index. 
\end{algorithmic}
\label{algorithm:spt}
\end{algorithm}

\section{Empirical findings}\label{Sec_Empirical findings}
In this section, we discuss the empirical evidence offered by the spatial and spatiotemporal clustering algorithms, motivating the choice of parameters and interpreting the results.

\subsection{Spatial Clustering} \label{Sec_Res_SpatCluster}
For the Spatial clustering, we considered the sample of 617 European firms with available ESG, Environmental and Carbon Emission scores for 2023. Recall that we obtain the dissimilarity matrices $D_0$ and $D_1$ from the Euclidean distances of the ESG-related variables and the geodetic distances of the coordinates of the firms, respectively.

\subsubsection{Choice of $K^*$ and $\alpha_K^*$}
\begin{figure}
\includegraphics[width=1\linewidth]{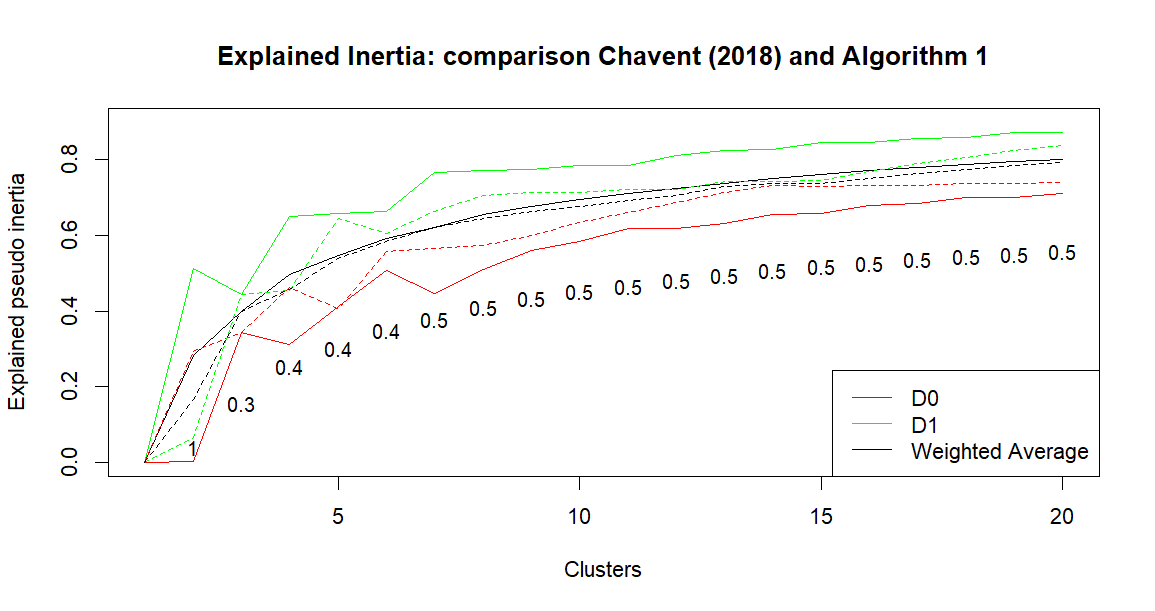}
\caption{Comparison of proportion of explained pseudo inertia from each dissimilarity matrices $D_0$ and $D_1$ and their weighted average, using \cite{Chavent2018} method (dashed line), and Algorithm \ref{algorithm:sp} (solid line) for $K=1,\ldots,K_{max}=20$. The values underlying the curves indicate the optimal $\alpha_K^*$ for Algorithm \ref{algorithm:sp}.}
\label{fig:sp_inertia}
\end{figure}

We carry out the spatial cluster analysis by setting $\Delta\alpha=0.1$ and $K_{max}=20$. We compute the optimal hyperparameters $\alpha^*$ and $K^*$ considering both the \cite{Chavent2018} procedure and the proposed Algorithm \ref{algorithm:sp}. As shown in Figure \ref{fig:sp_inertia}, our algorithm leads to a higher proportion of explained pseudo inertia in the spatial component and, as a consequence, the proportion of pseudo explained inertia is lower for the socio-economic features component. Overall, the weighted average proportion is higher than the proportion reached with the Chavent methodology, meaning that we are capturing better the overall variability embedded in the data. 

\begin{figure}
    \centering
    \begin{subfigure}{0.48\textwidth}
        \includegraphics[width=\linewidth]{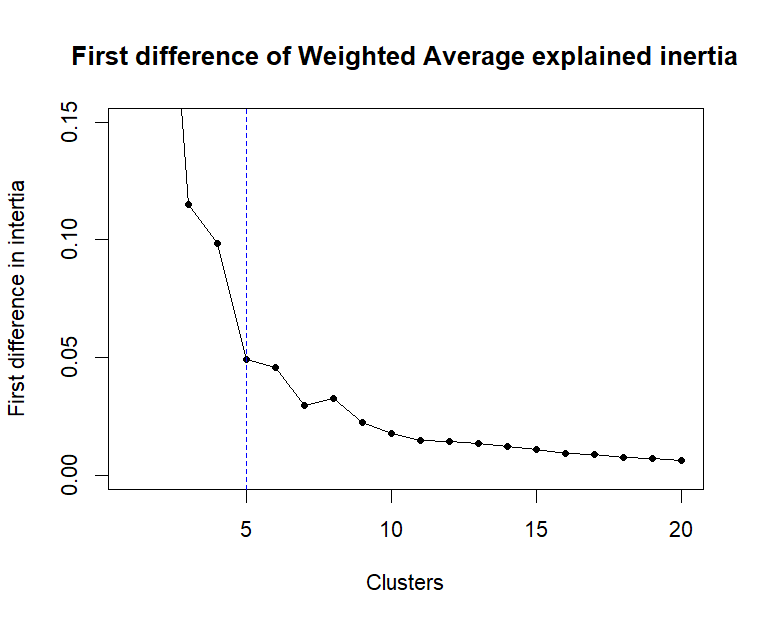}
        \caption{Gain in weighted average explained inertia}
    \end{subfigure}
    \hfill
    \begin{subfigure}{0.48\textwidth}
        \includegraphics[width=\linewidth]{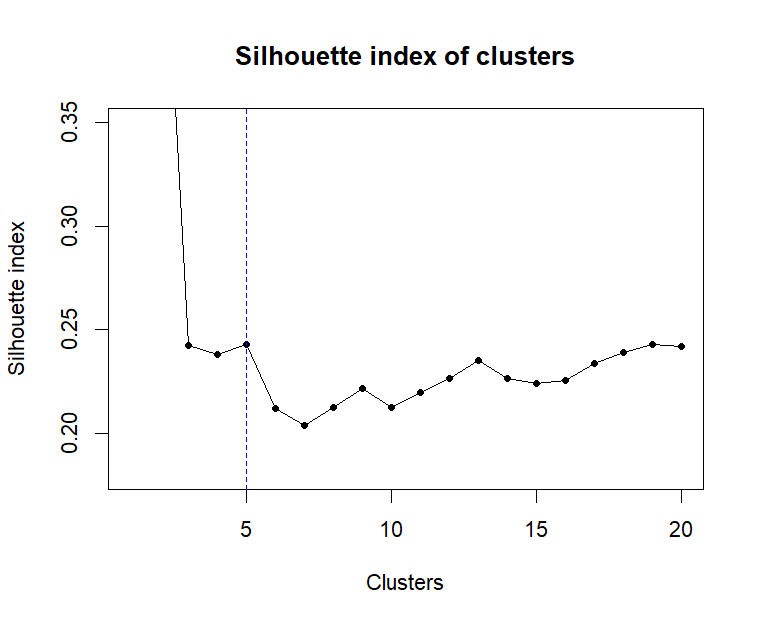}
        \caption{Silhouette index}
        \label{fig:sp_k}
    \end{subfigure}
    \caption{Hyperparameters selection in spatial clustering. Left panel: increment in the weighted average proportion of explained inertia generated by a unitary increase in the number of clusters. Right panel: Silhouette index computed for each partition into $K$ clusters. Recall that, for each $K$, we considered the best combination of the dissimilarity matrices according to the $\alpha_K^*$ found using the maximization inertia criterion according to Figure \ref{fig:sp_inertia}.}
\end{figure}

As regards the choice of the parameters, we consider the increment in the weighted average proportion of the explained inertia associated with a unitary increase in the number of clusters and the Silhouette index, where the dissimilarity matrices are linearly combined with respect to $\alpha_K^*$. The values of $\alpha_K^*$ are reported in Figure \ref{fig:sp_inertia}, while the aforementioned indices are described in Figure \ref{fig:sp_k}. We select $K^*=5$ because it seems a good compromise, able to gain a relevant increase in the weighted average explained inertia and a reasonable value of the Silhouette index. Consequently, we have $\alpha_{K=5}^*=0.40$.

For a more exhaustive comparison between the results obtained by the two methods, in Table \ref{tab:inertia} are reported the inertia, the proportion of explained inertia and the normalized proportion of the explained inertia by the two dissimilarity matrices, considering $K^*=5$ number of clusters. With respect to \cite{Chavent2018} approach, the proportion of explained inertia increases from $0.4070$ to $0.4120$ in the dissimilarity matrix $D_0$, and the gain in $D_1$ is from $0.6453$ to $0.6567$. Consequently, the normalized proportion of explained inertia obtained with $\alpha_5^*$ is higher for both matrices. 
\begin{table}[h]
    \centering
    \captionsetup{width=1\textwidth}
    \begin{tabular}{ll c lll c lll}
        \hline
        ~ & ~ && \multicolumn{3}{c}{Chavent $\alpha_{K=5}=0.50$} && \multicolumn{3}{c}{Algorithm \ref{algorithm:sp} $\alpha_{K=5}^*=0.40$}   \\ 
        \hline
        ~ & $W(\mathcal{P}_1)$ && $W(\mathcal{P}_K^{\alpha})$ & $\mathcal{Q}(\mathcal{P}_K^{\alpha})$ & $\tilde{\mathcal{Q}}(\mathcal{P}_K^{\alpha})$ && $W(\mathcal{P}_K^{\alpha})$ &  $\mathcal{Q}(\mathcal{P}_K^{\alpha})$ &  $\tilde{\mathcal{Q}}(\mathcal{P}_K^{\alpha})$  \\  
        \hline
        $D_0$ & 0.0473 && 0.0275 & 0.4070 & 0.5709 && 0.0271 & 0.4120 & 0.5780  \\ 
        $D_1$ & 0.0581 && 0.0221 & 0.6453 & 0.7802 && 0.0206 & 0.6567 & 0.7941  \\ 
        \hline
        \\
    \end{tabular}
    \caption{Summary of the inertia (absolute, relative and normalized) returned by the spatial clustering. Columns 3 to 5 report results according to \cite{Chavent2018}, that is, $K^*=5$ and $\alpha^*=0.50$. Columns 6 to 8 report results according to Algorithm \ref{algorithm:sp}, that is, $K^*=5$ and $\alpha^*=0.40$. $W(\mathcal{P}_1)$ is the total inertia from spatial dissimilarity (i.e., $W_{D_1}(\mathcal{P}_1)$) or socio-economic dissimilarity (i.e., $W_{D_0}(\mathcal{P}_1)$); $W(\mathcal{P}_K^{\alpha})$ is the absolute within-cluster pseudo inertia from the mixed clustering; $\mathcal{Q}(\mathcal{P}_K^{\alpha})$ is the proportion of explained pseudo-inertia; $\tilde{\mathcal{Q}}(\mathcal{P}_K^{\alpha})$ is the normalized proportion of inertia.}
    \label{tab:inertia}
\end{table}

\subsubsection{Resulting spatial Clusters}

\begin{figure}
\centerline{
\includegraphics[width=1.3\linewidth]{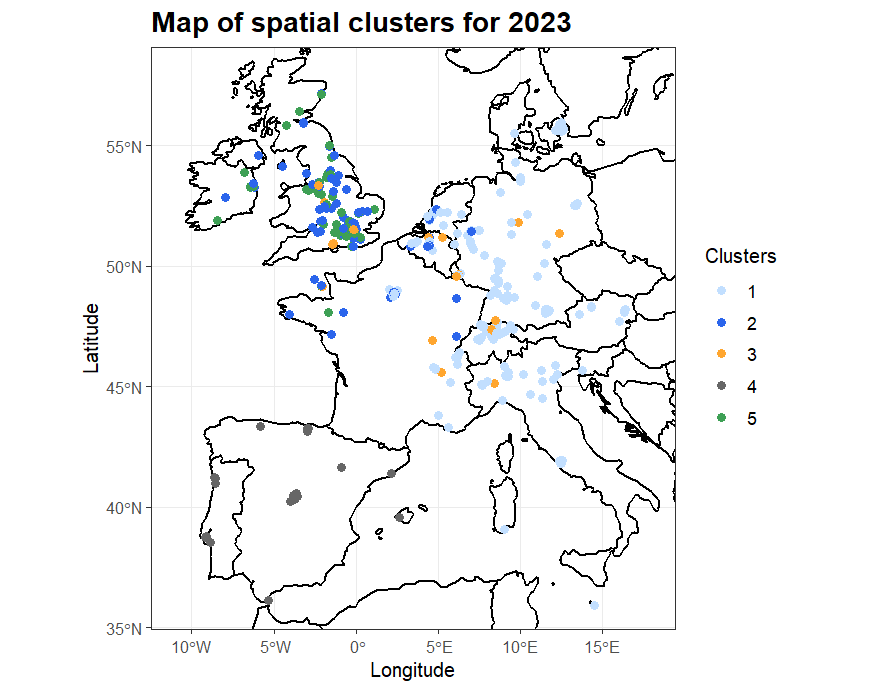}
}
\caption{Map of spatial clusters for 2023 using $K^*=5$ and $\alpha_5^*=0.40$. Clusters are computed using two dissimilarity matrices: Geodetic spatial distance and Euclidean distance of ESG, Environmental and Carbon Emission scores.}
\label{fig:sp_map}
\end{figure}

Figure \ref{fig:sp_map} maps the clusters produced by the spatial clustering for 2023 with parameters $K^*=5$ and $\alpha_5^*=0.40$. Each point represents a company location, with different colors depending on the cluster to which the company is assigned. Also, Figure \ref{fig:sp_composition} summarizes the distribution of firms by country (left panel) and by industry (right panel) to understand the obtained categorization better. 

In terms of number of companies and country representativeness, cluster 1 is definitely the largest group, with 273 companies located in 10 out of 15 countries in the sample, while the other clusters have fewer companies. Clusters 2 and 3 collect firms from 7 and 8 countries, respectively, while cluster 4 is composed only of companies from the Iberian Peninsula, and cluster 5 is mainly composed of UK and Irish companies (and one company from northern France). Moreover, cluster 3 and cluster 4 are the ones with the fewest companies, respectively 43 and 45. While cluster 4 includes all companies from Spain and Portugal (and one more from Gibraltar), the fact that cluster 3 is made up of companies from 8 countries suggests that there may be some common ESG factors that ultimately group these companies together. In summary, in terms of geographical overlap, cluster 1 seems to overlap with cluster 2 mainly in Belgium, the Netherlands, and France and to a minimal extent with other clusters. Instead, clusters 2, 3, and 5 show a high degree of overlapping in the UK and Ireland, while cluster 4 does not seem to overlap with other clusters.

\begin{figure}
\centering
\includegraphics[width=10cm]{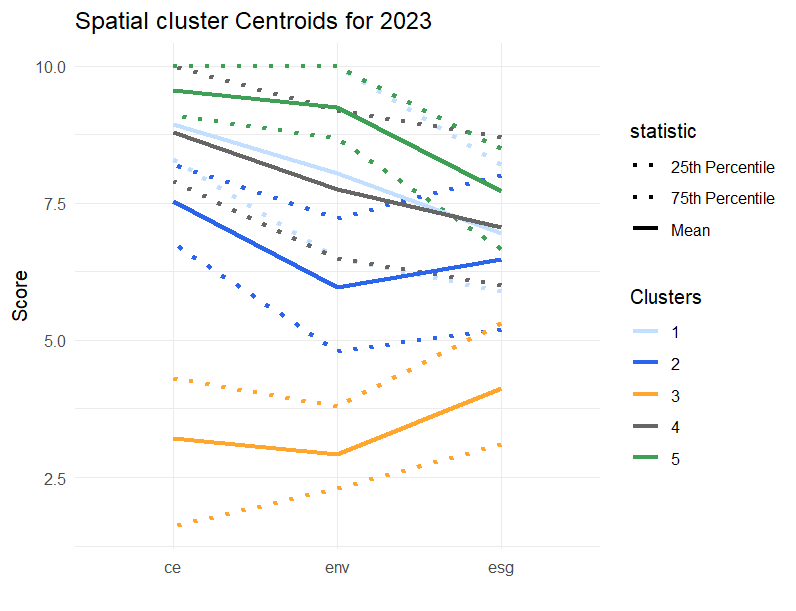}
\caption{Cluster-specific centroids of the three economic variables used to define the spatial clusters for 2023 using $K^*=5$ and $\alpha_5^*=0.40$. Recall that clusters are computed using two dissimilarity matrices: Geodetic spatial distance and Euclidean distance of ESG (esg), Environmental (env) and Carbon Emission (ce) scores. The solid lines represent the means of the variables, dashed lines represent the two quartiles.}
\label{fig:sp_centr}
\end{figure}

\begin{figure}
\centerline{
\includegraphics[width=1\linewidth]{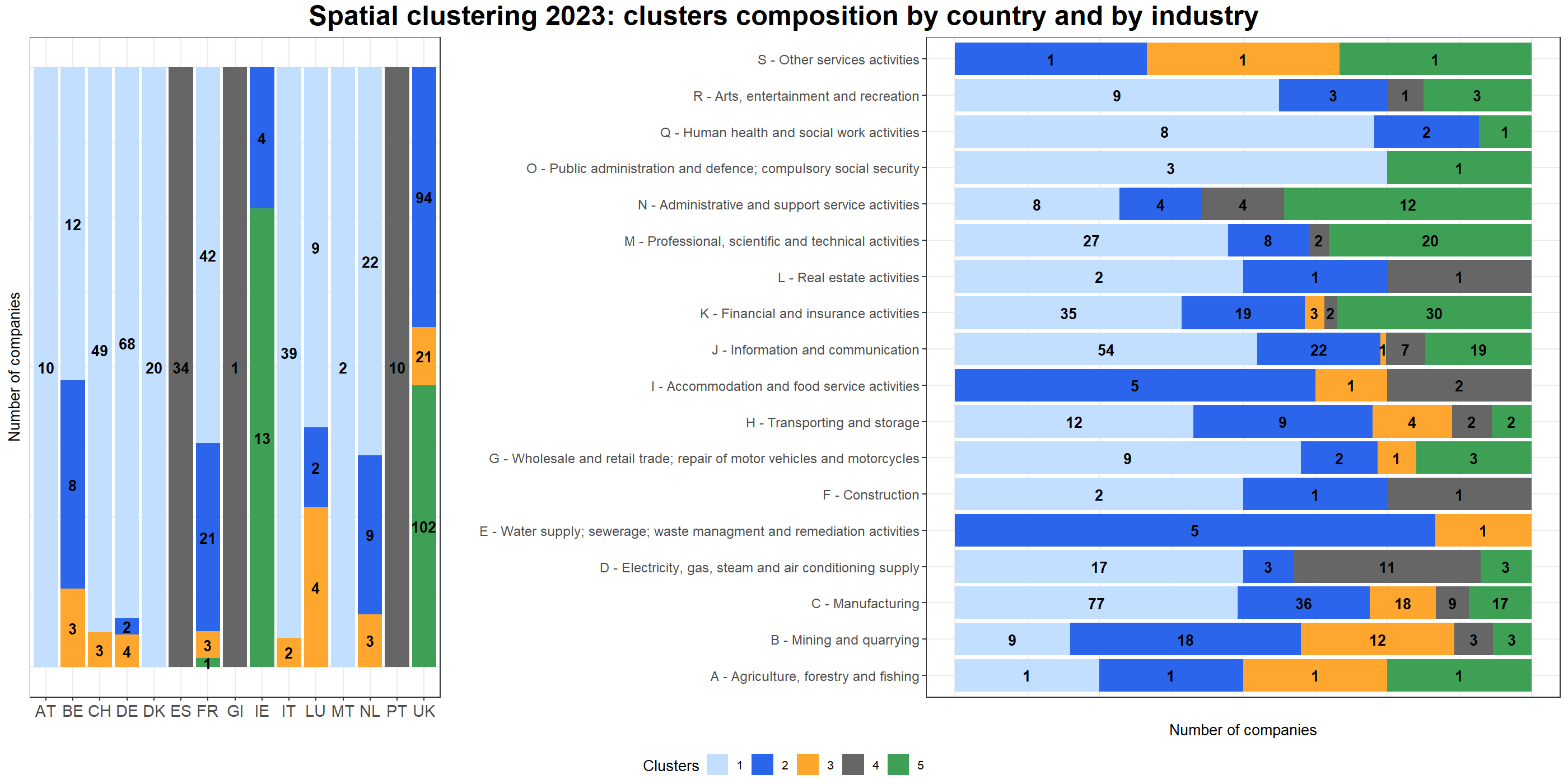}
}
\caption{Spatial clusters composition by country (specified using ISO code) and industry for 2023.}
\label{fig:sp_composition}
\end{figure}

For the environmental assessment of each cluster, Figure \ref{fig:sp_centr} shows the descriptive statistics (mean, 25th and 75th percentile range) of the clusters in terms of Carbon Emission score (ce), Environmental Pillar score (env), and ESG score (esg).

Clusters 1, 4, and 5 have similar patterns of sustainability scores; in particular, they record above-average Carbon Emission scores and Environmental scores. As their ESG score is, on average, lower than their Environmental Pillar score, the companies in this cluster have better environmental practices with respect to the Social and Governance Pillars. Cluster 5 seems to be the one with slightly larger mean scores than the other two. Cluster 2 shows a different pattern from the previous ones. Companies in this cluster seem to have higher carbon emissions than the environmental pillar scores, with their overall ESG score almost at the same level as Clusters 1 and 4. This indicates a disparity of ESG evaluation across clusters: companies in cluster 2 may lag in the overall environmental performance but exhibit relatively strong social and governance attributes, resulting in an ESG score comparable to other clusters, on average. 
Finally, cluster 3 seems to collect the companies with the lowest performance in terms of both carbon emissions score, environmental pillar score, and ESG score.  This finding is relevant given the geographical distribution of clusters 2 and 3. Despite their considerable overlap, our algorithm is able to differentiate companies based on their environmental performance relative to the other two pillars.

Although ESG scores are industry-based, examining the distribution of sectors in the different clusters we obtain from the model is particularly relevant. Considering the distribution across industries (based on the NACE industrial classification), Figure \ref{fig:sp_composition} shows a remarkable cross-sectorial heterogeneity. That is, there are no industries that are fully concentrated in a few clusters. However, the concentration of industries can vary considerably between clusters. For example, while in clusters 1, 2, and 3 the manufacturing industry (NACE C) is the most relevant sector (in both absolute and relative number of firms), in cluster 4 there are more enterprises in electricity, gas, steam, and air conditioning supply (NACE D), while in cluster 5 manufacturing enterprises are outnumbered by companies in information and communication (NACE J), financial and insurance activities (NACE K) and professional, scientific and technical activities (NACE M). It is worth noting that cluster 3, which is composed of enterprises with poor environmental sustainability performance, includes enterprises from ten NACE sectors, although six of them are represented by only one enterprise.

\subsection{Spatiotemporal Clustering}
Regarding spatiotemporal clustering application, starting from the initial sample of 1045 companies with at least one rating between 2013 and 2023, we considered the subsample of 460 firms with at least six annual observations within the same window. This choice is necessary to use the DTW distance for time series, which requires at least one overlapping time stamp for each pair of companies. Thus, having considered an 11-year period, the minimum number of overlapping years must be set to six; that is, for every selected company, we require a non-missing value for more than half of the entire period under consideration. We computed the dissimilarity matrices of the time series for ESG score, Environmental Pillar score and Carbon emission score, using the DTW algorithm described in Section \ref{Sec_Method_STclu}, and the geodetic distances from the coordinates. Thus, we obtained the dissimilarity matrices named $D_{esg},D_{env},D_{ce}$ and $D_{sp}$.

\subsubsection{Choice of $K^*$ and $\alpha_{Kp}^*$}
\begin{figure}
    \centering
    \includegraphics[width=0.85\linewidth]{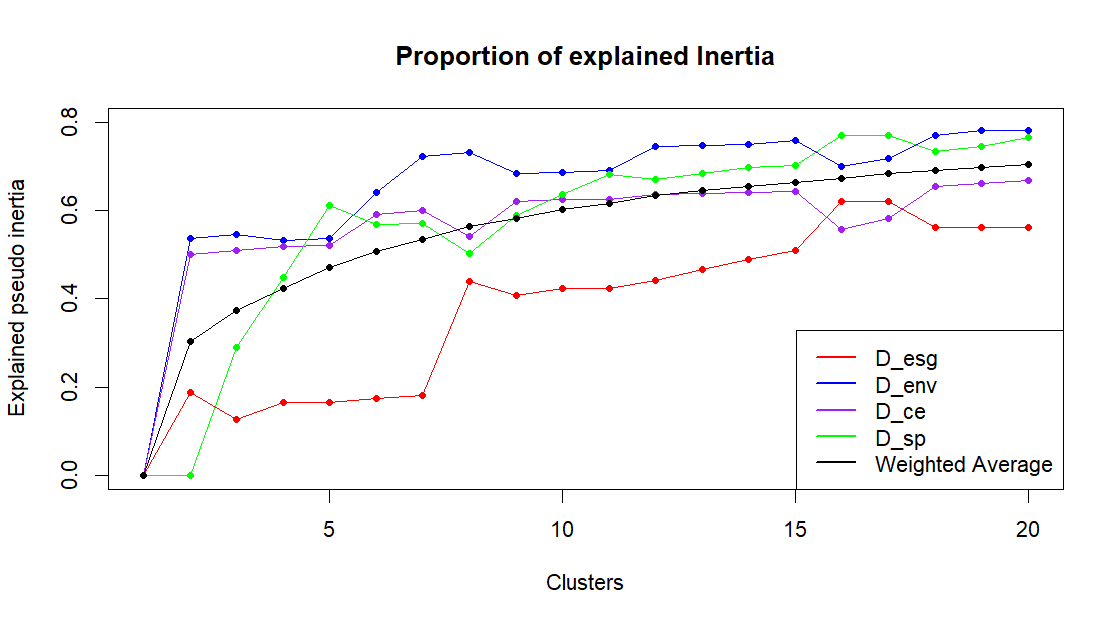}
    \includegraphics[width=0.85\linewidth]{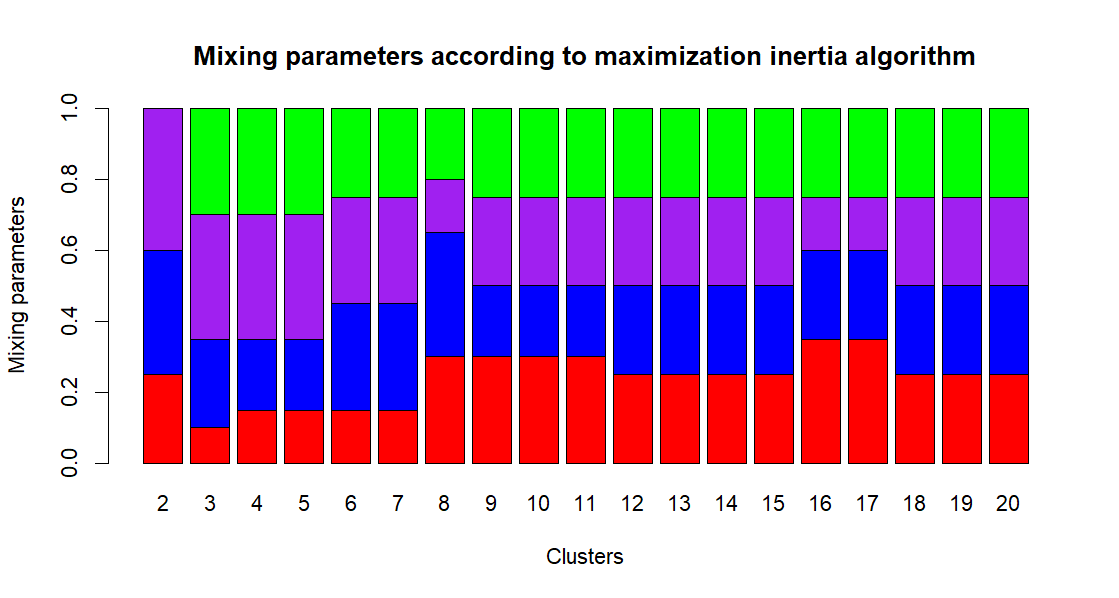}
    \caption{Top panel: proportion of explained pseudo inertia contained in each dissimilarity matrix (colored lines) and weighted average proportion (black line). Values are computed considering for each $K$ the optimal weighting combination $\alpha_{K,p}^*$ from Algorithm \ref{algorithm:spt}. Bottom panel: optimal weighting combination $\alpha_{K,p}^*$ from Algorithm \ref{algorithm:spt}. Colors refer to the four dissimilarity matrices used in the computation.}
    \label{fig:spt_alpha}
\end{figure}

To choose the optimal number of clusters $K$ and the linear combination of weights $\alpha_p$, we run the Algorithm \ref{algorithm:spt} setting $K_{max}=20$ and $\Delta \alpha=0.05$. Having to choose a higher number of mixing parameters, we think it is more appropriate to use a smaller $\Delta \alpha$, so as to consider a greater number of combinations, including the case when the matrices all have the same weight (i.e., $\alpha_p=0.25 \quad \forall p=1,\ldots,P$). Notice that the latter combination would not be considered if we used a $\Delta \alpha=0.1$ as in the purely spatial setting. In Figure \ref{fig:spt_alpha}, we show the results obtained for each number of clusters, both in terms of the proportion of explained inertia for each dissimilarity matrix and the weighted average proportion of explained inertia (top panel) and the best combination of mixing parameters.

\begin{figure}
    \centering
    \begin{subfigure}{0.48\textwidth}
        \includegraphics[width=\linewidth]{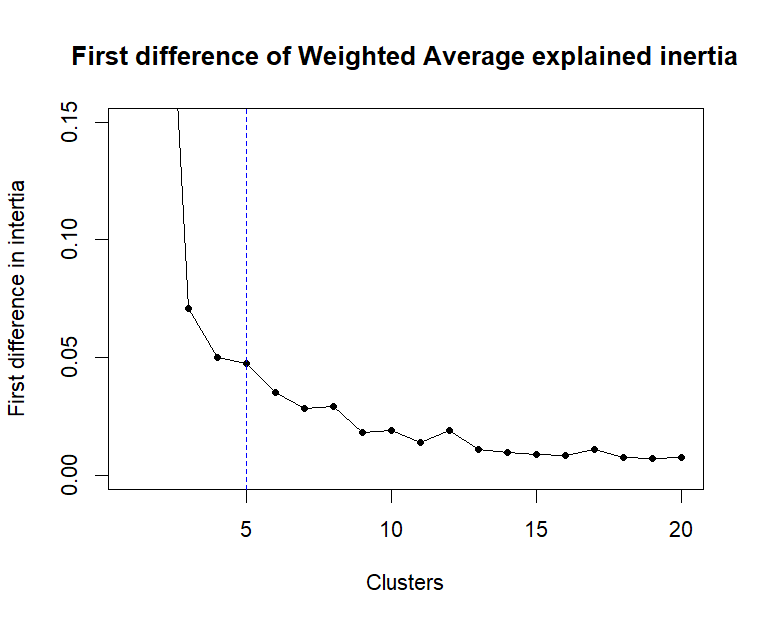}
        \caption{Gain in the weighted average of explained pseudo inertia }
    \end{subfigure}
    \hfill
    \begin{subfigure}{0.48\textwidth}
        \includegraphics[width=\linewidth]{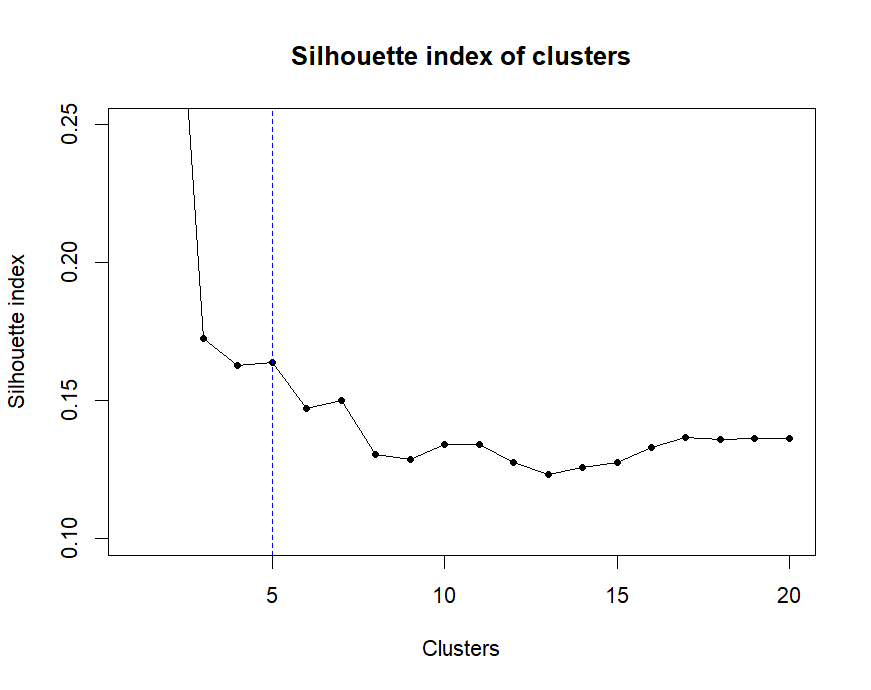}
        \caption{Silhouette index}
        \label{fig:spt_k}
    \end{subfigure}
    \caption{Hyperparameters selection in spatiotemporal clustering. Left panel: increment in the weighted average proportion of explained inertia generated by a unitary increase in the number of clusters. Right panel: Silhouette index computed for each partition into $K$ clusters. Recall that, for each $K$, we considered the best combination of the dissimilarity matrices according to the $\alpha_{Kp}^*$ found using the maximization inertia criterion according to Figure \ref{fig:spt_alpha}.}
\end{figure}

From the plots, it is possible to notice that the spatial component is included when considering at least $K=3$ groups. The dissimilarity matrix $D_{esg}$ plays a less important role for a number of clusters lower than 8. For a higher number of clusters, the best combinations of the dissimilarity matrices provide similar weights. In Figure \ref{fig:spt_alpha}, we compare the gain in the weighted average of explained inertia and the silhouette index from $K=1$ up to $K_{max}=20$, evaluating each $K$ at the corresponding optimal combination $\alpha_{K,p}^*$. Observing \ref{fig:spt_k}, when choosing a number of clusters equal to $K=5$, we manage to increase the weighted average proportion of inertia by 0.0478, and we get a silhouette index equal to 0.1640, which represents one of the highest values among those shown. Although the value of the silhouette may not seem too high, with the increasing dimensionality of the data, it becomes difficult to achieve high values because of the curse of dimensionality as the distances become more similar. 

Overall, we can state that using $K^*=5$ clusters with weights $\alpha_{5,esg}^*=0.15$, $\alpha_{5,env}^*=0.20$, $\alpha_{5,ce}^*=0.35$ and $\alpha_{5,sp}^*=0.30$ represent the best solution. In Table \ref{tab:inertia_spt}, we report the absolute explained inertia, the proportion of explained inertia and the normalized proportion of explained inertia for each dissimilarity matrix, considering our choice of hyperparameters. Overall, through cluster analysis, we are able to explain more than 50\% of inertia in the dissimilarity matrices $D_{env},D_{ce},D_{sp}$, but the proportion of explained inertia in $D_{esg}$ is lower than 25\%. 

\begin{table}[h]
    \centering
    \captionsetup{width=1\textwidth}
    \begin{tabular}{llllll}
    	\hline
        ~ & $\alpha_p$ & $W(\mathcal{P}_1)$ & $W(\mathcal{P}_K^{\alpha_p})$ & $\mathcal{Q}(\mathcal{P}_K^{\alpha_p})$ & $\tilde{\mathcal{Q}}(\mathcal{P}_K^{\alpha_p})$   \\ 
        \hline
        $D_{esg}$ & 0.15 & 0.0501 & 0.0418 & 0.1649 & 0.2195 \\ 
        $D_{env}$ & 0.20 & 0.0672 & 0.0311 & 0.5362 & 0.6411  \\
        $D_{ce}$ & 0.35 & 0.0456 & 0.0218 & 0.5206 &  0.6761\\ 
        $D_{sp}$ & 0.30 & 0.0626 & 0.0242 & 0.6125 &  0.7453 \\
        \hline
        \\
    \end{tabular}
\caption{
Summary of the inertia (absolute, relative and normalized) returned by the spatiotemporal clustering at the optimal solution $K^*=5$. $W(\mathcal{P}_1)$ is the total inertia provided by each dissimilarity matrix; $W(\mathcal{P}_K^{\alpha_p})$ is the absolute within-cluster pseudo inertia from the mixed clustering for each matrix; $\mathcal{Q}(\mathcal{P}_K^{\alpha_p})$ is the proportion of explained pseudo-inertia; $\tilde{\mathcal{Q}}(\mathcal{P}_K^{\alpha_p})$ is the normalized proportion of inertia.
}
\label{tab:inertia_spt}
\end{table}

In order to achieve a substantially higher proportion of explained inertia in $D_esg$, it is necessary to choose a much higher number of clusters, but this would complicate the interpretation of the resulting clusters. In addition, the silhouette index assumes even lower values when the number of clusters is greater than 5, so the homogeneity of the resulting clusters seems to decrease. Thus, $K^*=5$ is a reasonable choice in our spatiotemporal cluster analysis.

\subsubsection{Resulting Spatiotemporal Clusters}
\begin{figure}
\centerline{
\includegraphics[width=1.3\linewidth]{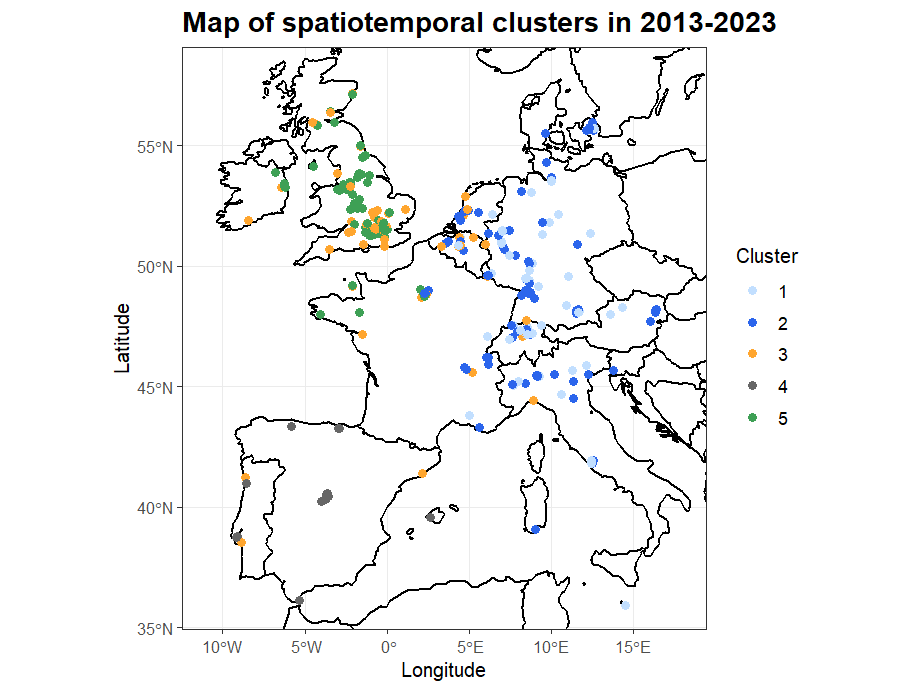}
}
\caption{Map of spatiotemporal clusters between 2013 and 2023. Clusters are computed using four dissimilarity matrices: Geodetic spatial distance, DTW distance of the overall ESG scores, DTW distance of the Environmental scores and DTW distance of Carbon Emission scores.}
\label{fig:spt_map}
\end{figure}

The spatiotemporal clustering produces substantially different groups with respect to the spatial clustering results described in Section \ref{Sec_Res_SpatCluster}. Figure \ref{fig:spt_map} represents the companies based on their location and on the final cluster to which they belong.

It is possible to notice the higher degree of overlap between the five clusters with respect to the spatial clustering analysis. Even in the Iberian Peninsula, we observe the presence of two clusters, namely 3 and 4, whereas in the previous analysis, only one cluster was present. This implies that the specific temporal dynamics of ESG scores could be relevant in our multidimensional approach. In other words, the inclusion of the temporal component captures new information that was missed when only the spatial component was considered. In particular, different pairs (or sometimes even triplets) of clusters could be identified in different geographical areas. Clusters 1 and 2 overlap in Italy, Denmark and Germany, and also with some enterprises from cluster 3 in Switzerland; clusters 2 and 3 overlap mainly in Belgium and the Netherlands, and they are also found together with cluster 5 in Paris; clusters 3 and 5 overlap significantly in the UK and Ireland; in Portugal, there is an overlap between clusters 3 and 4, especially around the cities of Lisbon and Porto. For the environmental and sustainability assessment, we represent the mean, the 25th and the 75th percentile of the Carbon Emission Score, Environmental Pillar score and ESG score in the period observed for the different clusters.

A synthesis of the results by country and by industry is reported in Figure \ref{fig:sp_composition}. Similarly to the previous spatial analysis, the distribution of sectors across clusters is quite heterogeneous (Figure 11). Two results stand out. Once again, the manufacturing sector (NACE C) is not dominant in all clusters, but only in clusters 1, 2 and 3. In cluster 4, there are more enterprises in electricity, gas, steam and air conditioning supply (NACE D), information and communication (NACE J) and administrative and support service activities (NACE N) than in manufacturing. In cluster 5, on the other hand, manufacturing is outnumbered by information and communication (NACE J) and finance and insurance (NACE K). Finally, cluster 3, the worst-performing cluster in terms of environmental sustainability, includes many manufacturing enterprises and most enterprises in mining and quarrying (NACE B) and transport and storage (NACE H). 
\begin{figure}
\centerline{
\includegraphics[width=1\linewidth]{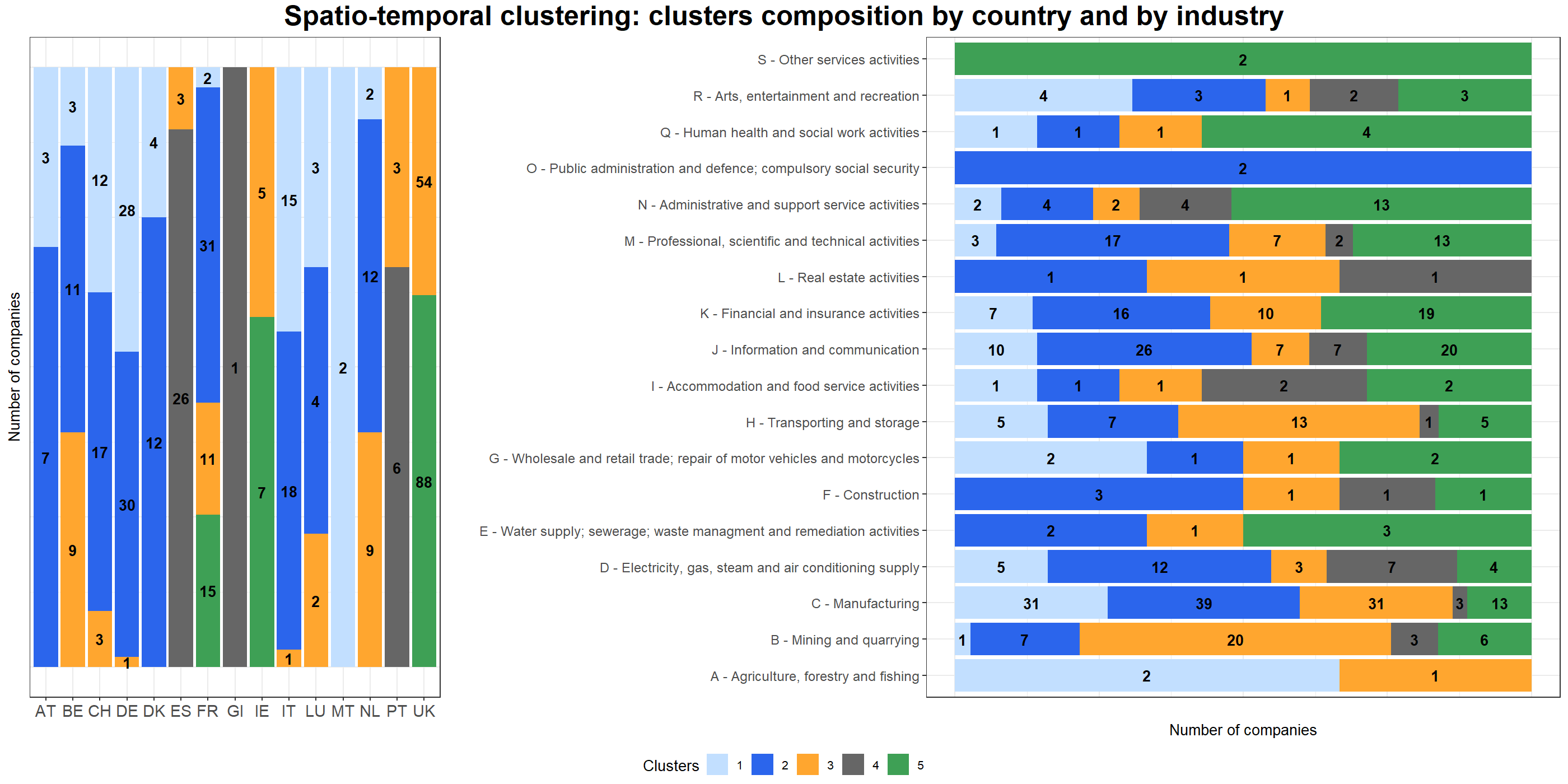}
}
\caption{Spatiotemporal clusters (from 2013 to 2023) composition by country (specified using ISO code) and industry.}
\label{fig:spt_composition}
\end{figure}

As in the spatial clustering analysis, some clear paths can be identified in the individual clusters, as well as some common aspects between the clusters.
\begin{figure}
\centering
\includegraphics[width=1\linewidth]{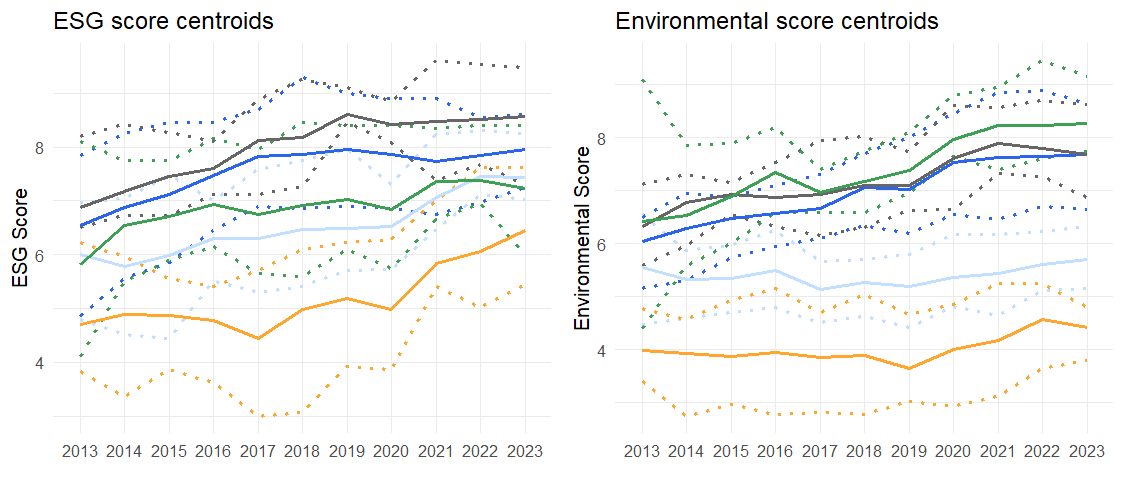}
\includegraphics[width=0.68\linewidth]{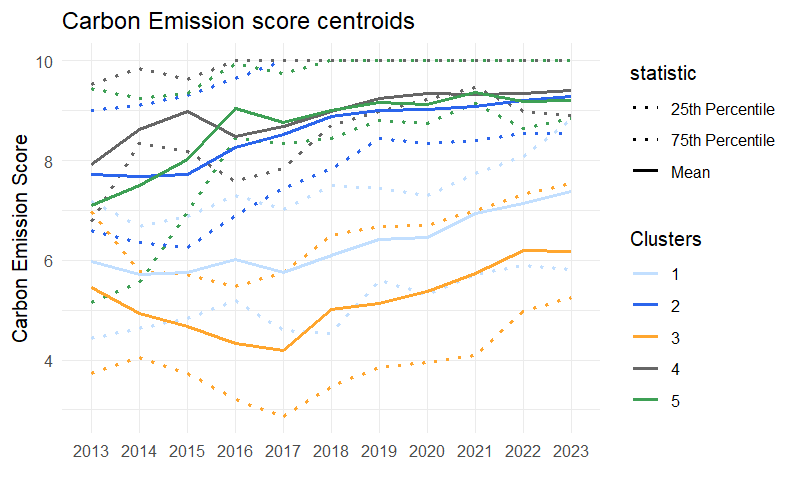}
\caption{Time series 2013-2023 of centroids (solid lines) and quantiles (dashed lines) of the socio-economic used for the spatiotemporal clusters.}
\label{fig:spt_centr}
\end{figure}
As reported in Figure \ref{fig:spt_centr}, clusters 2, 4 and 5 seem to contain the companies with the highest performance in all three variables considered. Their scores also seem to increase over time, albeit with different magnitudes. On the contrary, with respect to the previous clusters, clusters 1 and 3 collect companies with visibly lower levels of environmental performance and carbon emission scores, with cluster 3 being the lowest-scoring cluster on average. While the pattern for the ESG score for the two clusters appears to be increasing throughout the period, as for the other three clusters, the pattern for the environmental score appears to be constant throughout the period. Instead, the two clusters observe a very different path for the carbon emission score. For cluster 3, this variable decreases on average between 2013 and 2017, while after 2017, it seems to increase until the end of the period. For cluster 2, conversely, the carbon emission score seems to be constant in 2013-2017 and then starts to increase until 2023.


\section{Conclusion}\label{Sec_Conclusions}

In this paper, we used spatial and spatiotemporal clustering methodologies to identify the main patterns that characterize the environmental performance of European companies. In particular, we examined the role of spatial location and temporal dynamics for European companies with ESG scores in the period between 2013 and 2023. For this reason, we developed a spatial and a spatiotemporal version of a hierarchical clustering technique accounting for multiple dimensions and providing alternative criteria to efficiently determine suitable values for the clustering hyperparameters, that is, the weighting combination between the considered variables and the optimal number of clusters.

Our findings suggest that both space and time matter when analyzing patterns of ESG performance. The spatial analysis for 2023 provided evidence of the presence of cross-national and cross-industrial groups of companies with remarkable differences in the levels of environmental performance. In particular, we were able to detect a subsample of companies with very poor environmental and ESG scores, which belong to several European countries and are mainly classified in the manufacturing and mining industries. Other clusters are mainly driven by companies engaged in the tertiary and service sectors, with a more regional and less transnational character (e.g., the clusters in the UK and in the Iberian peninsula), with higher levels of sustainability performance. Regarding the space-time dynamic, the identified groups are more prone to spatial overlapping (e.g., two clusters in the Iberian peninsula), suggesting that the ESG scores' temporal aspect is relevant to our multidimensional approach. As for the purely spatial analysis, also in the spatiotemporal setting, the dualism between the manufacturing sector and tertiary sectors is a key element, with the manufacturers dominating the worst environmental-performing clusters.

Finally, another outcome of interest is the gap (in some groups, considerable gaps exist) between overall ESG score values and the scores in the environmental and carbon emissions pillars. In this sense, both approaches show a distance between environmental scores with respect to the social and governance pillars. In other words,  companies with better environmental and emissions scores perform worse on average in the social and governance categories, lowering the overall ESG score. In contrast, companies with low environmental scores perform better in terms of social and governance. This difference could be due to divergence in corporate strategies, leading to specialization in internal company practices as well, and thus diverse scores.

In summary, the main take-home message of this paper is that firms' geographical locations are relevant for a comprehensive understanding of the time dynamics of ESG scores, particularly in explaining firms' ability to achieve positive scores, especially in the environmental aspects of sustainability. Conversely, the transnational nature of groups, i.e., the high degree of overlap between clusters, may pose a challenge when attempting to link cluster ESG performance to different national or supranational policies that influence firms' pursuit of sustainability practices.




\section*{Appendix}

\section{Further details on the average proportion of explained mixed pseudo inertia $\bar{Q}(\mathcal{P}_K^{\alpha})$} \label{App_DerivGain}
Let us consider a partition $\mathcal{P}_K^{\alpha}$ in $K$ clusters obtained by mixing the dissimilarity matrices $D_0$ and $D_1$ with the coefficient $\alpha$. Also, let us denote its within-clusters mixed inertia as $W(\mathcal{P}_K^{\alpha})$ and the corresponding proportion of the total pseudo inertia explained as
\begin{equation*}
Q_{0}(\mathcal{P}_K^{\alpha})=1-\frac{W_{0}(\mathcal{P}_K^{\alpha})}{W_{0}(\mathcal{P}_1)} \qquad Q_{1}(\mathcal{P}_K^{\alpha})=1-\frac{W_{1}(\mathcal{P}_K^{\alpha})}{W_{1}(\mathcal{P}_1)},
\end{equation*}
where $W_{0}(\mathcal{P}_1)$ and $W_{1}(\mathcal{P}_1)$ are the total pseudo inertia under dissimilarity matrix $D_0$ and under dissimilarity matrix $D_1$, respectively.

The two previous expressions can be reformulated as follows:
\begin{equation*}
Q_{0}(\mathcal{P}_K^{\alpha}) \cdot W_{0}(\mathcal{P}_1) = W_{D_0}(\mathcal{P}_1) - W_{D_0}(\mathcal{P}_K^{\alpha})
\end{equation*}
\begin{equation*}
Q_{1}(\mathcal{P}_K^{\alpha}) \cdot W_{1}(\mathcal{P}_1) = W_{D_1}(\mathcal{P}_1) - W_{D_1}(\mathcal{P}_K^{\alpha})
\end{equation*}

Now, let us denote the \textit{weighted average of explained mixed pseudo inertia for partition $\mathcal{P}_K^{\alpha}$} as the following linear combination:
\begin{equation*}
    \bar{Q}(\mathcal{P}_K^{\alpha}) = \frac{Q_{D_0}(\mathcal{P}_K^{\alpha}) \cdot W_{D_0}(\mathcal{P}_1) + Q_{D_1}(\mathcal{P}_K^{\alpha}) \cdot W_{D_1}(\mathcal{P}_1)}{W_{D_0}(\mathcal{P}_1) + W_{D_1}(\mathcal{P}_1)}.
\end{equation*}
In practice, we linearly combine the proportion of pseudo inertia explained by partition $\mathcal{P}_K^{\alpha}$ under dissimilarity $D_0$ (i.e., $Q_{D_0}(\mathcal{P}_K^{\alpha})$) and the share of pseudo inertia explained under dissimilarity $D_1$ (i.e., $Q_{D_1}(\mathcal{P}_K^{\alpha})$) weighting the values with the total pseudo inertia provided by the geographical information (i.e., $W_{D_0}(\mathcal{P}_1)$) and the total pseudo inertia provided by the socio-economic features (i.e., $W_{D_1}(\mathcal{P}_1)$).

Alternatively, the weighted average can also be expressed as a function of the total relative gain obtained by implementing a mixed partitioning instead of a purely-spatial or purely-socio-economic information approach, that is,
\begin{equation*}
    \bar{Q}(\mathcal{P}_K^{\alpha}) = \frac{[W_{D_0}(\mathcal{P}_1) - W_{D_0}(\mathcal{P}_K^{\alpha})] + [W_{D_1}(\mathcal{P}_1) - W_{D_1}(\mathcal{P}_K^{\alpha})]}{W_{D_0}(\mathcal{P}_1) + W_{D_1}(\mathcal{P}_1)}
\end{equation*}
where $W_{D_0}(\mathcal{P}_1) - W_{D_0}(\mathcal{P}_K^{\alpha})$ is the discrepancy between the overall pseudo inertia using $D_0$ only and the pseudo inertia induced by partition $\mathcal{P}_K^{\alpha}$ in $K$ clusters (i.e., the gain in inertia obtained by using the combination of the two matrices instead of the $D_0$ matrix alone), while $W_{D_1}(\mathcal{P}_1) - W_{D_1}(\mathcal{P}_K^{\alpha})$ is the discrepancy between the overall pseudo inertia using $D_0$ only and the pseudo inertia induced by partition $\mathcal{P}_K^{\alpha}$ in $K$ clusters (i.e., the gain in inertia obtained by using the combination of the two matrices instead of the $D_1$ matrix alone). This expression allows a further interpretation of the proposed criterion for selecting $\alpha$; that is, we are maximizing the relative gain in terms of inertia induced by using a mixture of geographical and socio-economic information to cluster the observations instead of employing a purely spatial or purely socio-economic clustering.

Eventually, by rearranging the previous expression, one can rewrite $\bar{Q}(\mathcal{P}_K^{\alpha})$ as a function of the within-cluster pseudo inertia as follows:
\begin{equation*}
\begin{split}
    \bar{Q}(\mathcal{P}_K^{\alpha}) &= \frac{W_{D_0}(\mathcal{P}_1) + W_{D_1}(\mathcal{P}_1) - \left[W_{D_0}(\mathcal{P}_K^{\alpha}) + W_{D_1}(\mathcal{P}_K^{\alpha})\right]}{W_{D_0}(\mathcal{P}_1) + W_{D_1}(\mathcal{P}_1)} \\
    &= 1 - \frac{W_{D_0}(\mathcal{P}_K^{\alpha}) + W_{D_1}(\mathcal{P}_K^{\alpha})}{W_{D_0}(\mathcal{P}_1) + W_{D_1}(\mathcal{P}_1)}.
\end{split}
\end{equation*}
Notice that the last expression allows for an easy generalization to the case of $p=1,2,\ldots,P$ dissimilarity matrices $D_1,\ldots,D_p,\ldots,D_P$, as in the case of spatiotemporal clustering with multiple dimensions,
\begin{equation*}
\bar{Q}(\mathcal{P}_K^{\alpha_p})=1-\frac{\sum_{p=1}^P W_{D_p}(\mathcal{P}_K^{\alpha_p})}{\sum_{p=1}^P W_{D_p}(\mathcal{P}_1^{\alpha_p})}.    
\end{equation*}


\end{document}